# Comparative Evaluation of Memory Technologies for Synaptic Crossbar Arrays - Part I: Robustness-driven Device-Circuit Co-Design and System Implications

Chunguang Wang*, Jeffry Victor*, and Sumeet K. Gupta, *Senior Member, IEEE*

***Abstract*—In-memory computing (IMC) utilizing synaptic crossbar arrays is promising for energy-efficient deep neural network (DNN) accelerators. Various technologies (CMOS and post-CMOS) have been explored as synaptic device candidates, each with its own pros and cons. In this work, we perform a design space exploration and comparative evaluation of four technologies viz. 8T SRAMs, ferroelectric transistors (FeFETs), resistive RAMs (ReRAMs) and spin-orbit torque magnetic RAMs (SOT-MRAMs) in the context of IMC robustness and DNN accuracy. For a fair comparison, we carefully optimize each technology specifically for synaptic crossbar design accounting for device and circuit non-idealities. By integrating different technologies into a cross-layer simulation flow based on physical models of synaptic devices and interconnects, we present insights into various device-circuit interactions. Based on the optimized designs, we obtain inference results for ResNet-20 on CIFAR-10 dataset. Among the four technologies, we show that FeFETs-based DNNs achieve the highest accuracy (followed closely by ReRAMs) and the largest resilience to process variations due to the compactness and high ON/OFF current ratio of FeFET bit-cells. In Part II of this paper, we expand the technology evaluation considering various device-circuit design knobs used for crossbar arrays.**

## I. INTRODUCTION

Deep neural networks (DNNs) have shown great promise for various machine learning applications such as computer vision, autonomous vehicles and others. However, suffering from the memory bottleneck and large memory footprints, CMOS-based von Neumann architectures are not able to provide adequate energy efficiency and on-chip memory density to meet the surging storage and computation demands of DNNs. In-memory computing (IMC) based on synaptic crossbar arrays is a promising alternative to alleviate expensive data transfers between memory and processor. Crossbar memory arrays not only store the DNN weights but also perform in-situ matrix-vector multiplication (MVM) which is the dominant kernel in DNNs. Further, to mitigate the low integration density of static random-access memories (SRAMs) [1], emerging non-volatile memories (NVMs) such as resistive RAMs (ReRAMs) [2], ferroelectric transistors (FeFETs) [3], spin-orbit torque magnetic RAMs (SOT-MRAMs) [4] and others are being explored as synaptic devices.

However, each memory technology has its own pros and cons, which needs to be properly understood. Several previous works [2], [4] have optimized and evaluated individual technologies in the context of synaptic crossbar array design. The authors in [5] perform a comparison of various emerging technologies for synaptic crossbar array designs, but focus on energy-latency-area analysis. However, another important design criterion for IMC designs is to ensure sufficiently large computational robustness in the presence of device-circuit non-idealities. Though several frameworks [6], [7], [8] have explored the impact of non-idealities in crossbar arrays, a detailed *comparative* analysis of different technologies and an understanding of the interaction of their unique device features with the non-ideal circuit attributes are lacking.

In this work, we aim to close this gap by utilizing a cross-layer simulation framework to capture device-circuit interactions and facilitate the connection between the technology attributes and DNN accuracy. We conduct a comparative evaluation of 8T-SRAMs, ReRAMs, FeFETs and SOT-MRAMs by analyzing their implications on IMC robustness and inference accuracy. In Part I of this paper, we focus on establishing IMC-driven device-circuit co-optimization strategies and the system implications of the four technologies. In Part II, we extend our comparative technology evaluation by considering various design knobs utilized for crossbar arrays. The key contributions of this part of the paper are as follows:

- Based on physical models for ReRAMs, FeFETs, SOT-MRAMs, 8T-SRAMs and interconnects, we explore the design space of synaptic crossbar arrays at the 7nm technology node and establish the optimal design choices to mitigate the impact of hardware non-idealities.
- Based on the optimized designs, we *comparatively* evaluate the four technologies by analyzing the accuracy of ResNet-20 on CIFAR-10 dataset.
- We provide insights into the device-circuit interactions and discuss the factors that dictate the performance of a technology for DNN inference.

This work was supported in part by the Center for Brain-Inspired Computing (C-BRIC), and in part by the Center for the Co-Design of Cognitive Systems (COCOSYS), funded by Semiconductor Research Corporation (SRC) and DARPA under Grant AWD-004311-S4.
(Corresponding author: Chunguang Wang.)

Chunguang Wang, Jeffry Victor, and Sumeet K. Gupta are with the Elmore Family School of Electrical and Computer Engineering, Purdue University, West Lafayette, IN 47907 USA (e-mail: wang4015@purdue.edu; louis8@purdue.edu; guptask@purdue.edu).

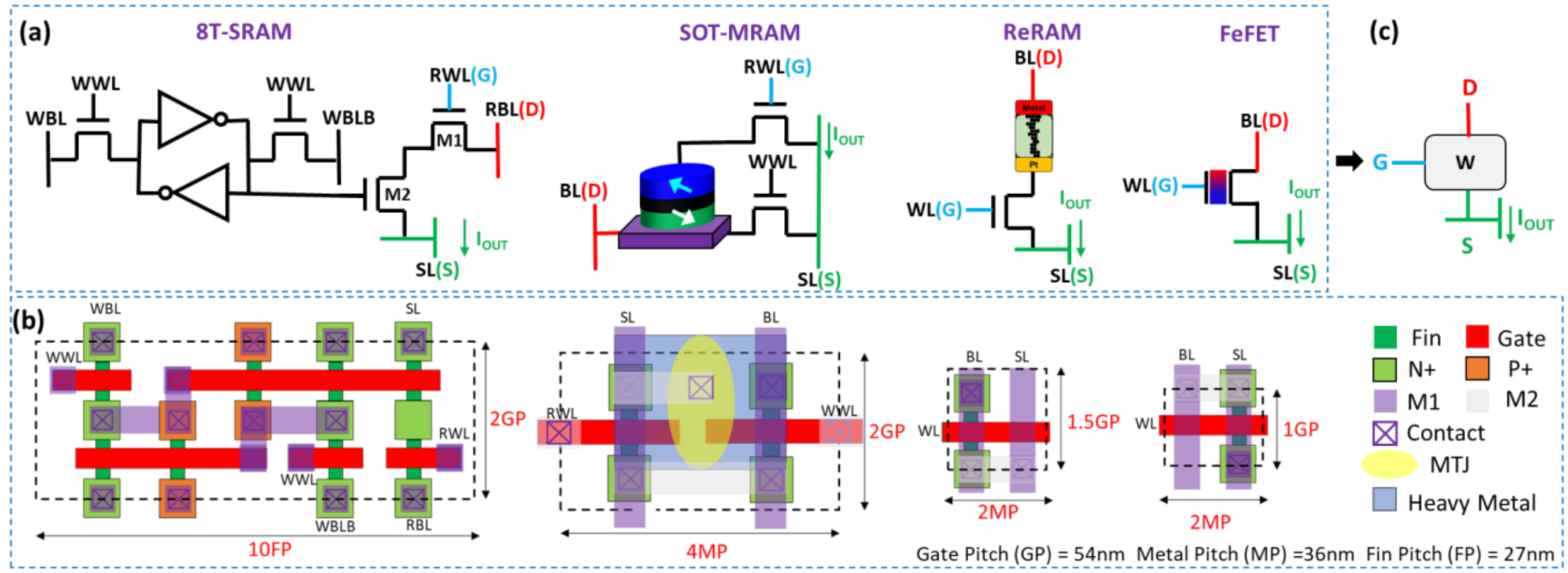

Fig. 1. (a) Schematic and (b) layout of bit-cell of 8T-SRAMs, SOT-MRAMs, ReRAMs and FeFETs. (c) Abstracted bit-cell with three terminals.

## II. Background and Related Works

The key attributes of IMC of MVM in a crossbar array for DNNs are: (a) synaptic weight encoding in the memory elements in the form of their conductance (b) input vector encoding as voltages applied on wordlines (WL) or bitlines (BL) running along the rows and (c) dot product outputs sensed at the sense-lines (SL) running along the columns. For the input voltage generation, especially for multi-bit input, digital-to-analog converters (DAC) can be used. Similarly, at the SLs, analog-to-digital converters (ADC) are typically needed to obtain a digital partial sum from the output current or the associated quantities (voltage or charge). The synaptic weights are encoded by programming the memory elements to binary or multiple levels to store a bit-slice of the weight. Similarly, the input bit-stream is applied on the WL or BL as binary or multi-level voltages.

While promising to obtain massively parallel in-situ computation of MVM, the actual hardware implementations of the crossbar arrays show a notable deviation from the ideal or expected output [2]. Such hardware non-idealities stem from device non-linearities and variability, finite driveability (i.e. non-zero resistance) of the word-line drivers, load resistance of the sensing circuitry, wire resistance ($R_{WIRE}$) and others [9]. As a result, the SL current exhibits non-linear behavior with respect to the output that it encodes and can be a source of sensing errors. This, in turn, leads to erroneous computations, which, if critical, can reduce the DNN accuracy. Hence, the optimization of the synaptic devices and crossbar arrays must account for such non-ideal behavior to minimize its impact on the inference accuracy.

Several previous works have analyzed the impact of non-idealities in crossbar arrays. NeuroSim [6] accounts for the impact of non-ideal device properties based on synaptic array simulations but does not comprehensively account for all device-circuit non-idealities. CxDNN [7] models the effects of parasitic resistances using matrix inversion techniques. However, non-linear non-idealities are not considered in that work. The works in [8] explore the fundamental limits on the compute signal-to-noise ratio (SNR) of crossbars by using statistical signal and noise models but the impact of technology-specific design parameters is not considered comprehensively. GENIEx [2] can capture the data-dependent non-idealities. However, sensing robustness and device variations have not been considered in the analysis. Further, in the existing works, there is a lack of a comparative evaluation of synaptic technology candidates considering the optimal device-circuit design choices that target minimizing hardware non-idealities at deeply scaled technology nodes. In this work, we carry out an extensive analysis of the device-circuit interactions accounting for the non-ideal technological and crossbar array attributes and perform a *comparative* technology evaluation in the context of IMC robustness and DNN accuracy at the 7nm technology node.

## III. Synaptic Crossbar Array Modeling

Based on the Arizona State Predictive process design kit (ASAP-7) [10] and the corresponding design rules (validated with the 7nm industrial process parameters such as gate/metal/fin pitch), we design the bit-cells (including their layouts) and the synaptic arrays for different technologies at 7nm technology node (Fig. 1). For IMC of MVM in a crossbar array, the read port of the bit-cells corresponding to different technologies can be abstracted as cells with three terminals: G, D and S, as shown in Fig. 1. FinFETs with single fin are used in the crossbar array for high-density and low capacitance and are modeled by the 7nm predictive technology multi-gate model (PTM-MG). For 8T-SRAMs, we employ the low standby power model for FinFETs to minimize the leakage. For the NVM technologies (which do not suffer from standby leakage issues), we utilize the high-performance FinFET model to lower the access transistor resistance.

We use 16-bit weights which are stored in 16 crossbar arrays, each storing a bit-slice of 1. We utilize multiple 64x64 crossbar sub-arrays for MVM in this part. In Part II, we conduct the analysis for different array sizes. Note, while SRAMs and SOT-MRAMs are limited to a bit-slice of 1, ReRAMs and FeFETs offer multi-level storage, which we analyze in Part II. In this part, we focus on a comparative analysis with a bit-slice of 1. We use 16-bit inputs which are streamed in 16 cycles as binary WL or BL voltages.

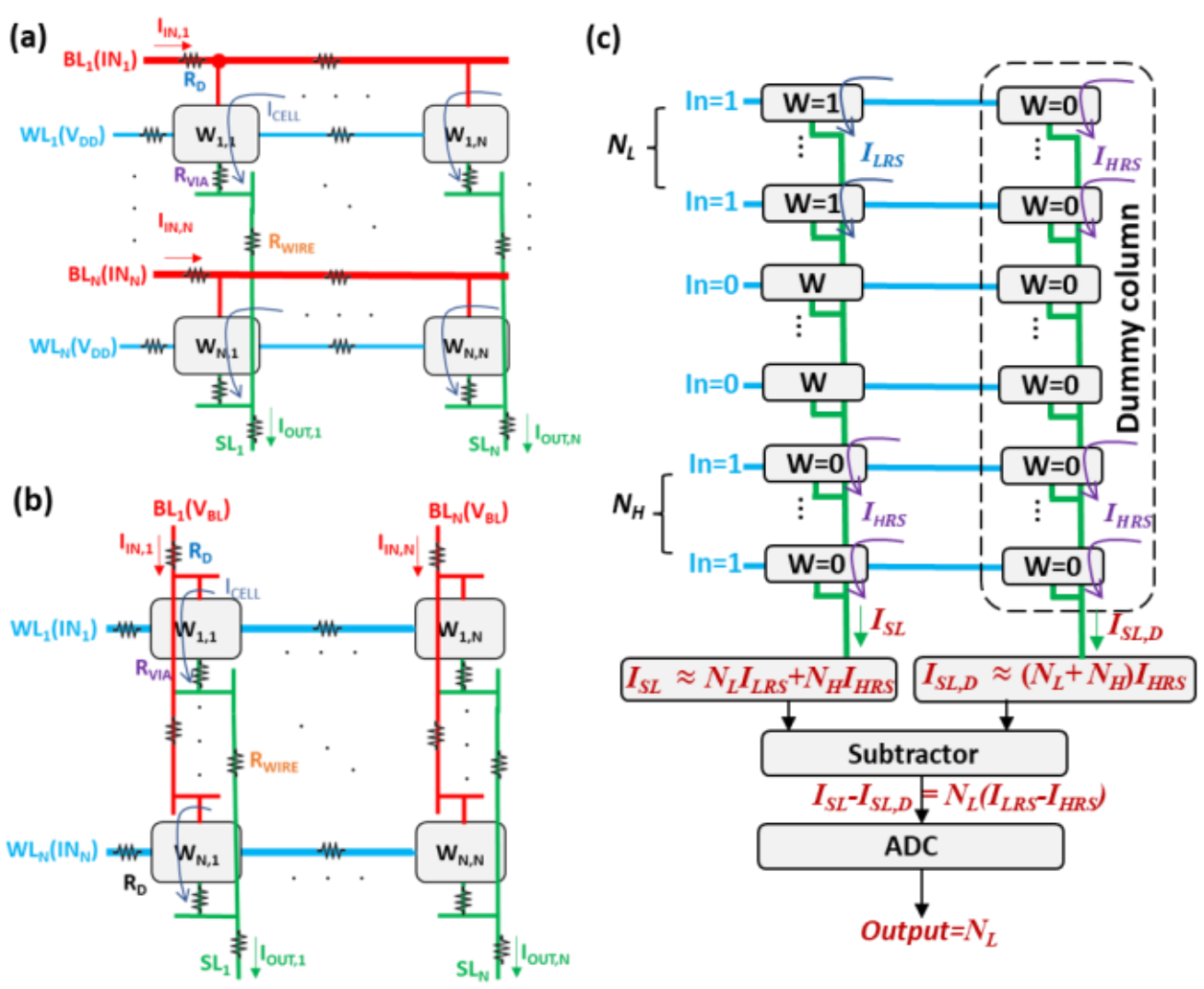

Fig. 2. (a) D-input configuration (b) G-input configuration of synaptic crossbar array. (c) A dummy column is included in the crossbar array to mitigate the impact of $I_{HRS}$.

We consider two options for input encoding: (1) D-input and (2) G-input. In D-input (Fig. 2 (a)), the inputs are applied to bitlines (BLs) which run along the row. For IMC, BLs corresponding to input = 1 are asserted simultaneously while all WLs are biased at $V_{DD}$ (=0.7V). D-input configuration is efficient for implementing multi-bits in the input stream. In the G-input configuration (Fig. 2 (b)), the inputs are applied on WLs. In this case, both D and S terminals of the bit-cells (connected to BL and SL, respectively) run along the column and all the BLs are biased at a fixed voltage (=0.25V in our analysis).

We also include a dummy column with each crossbar array [11], [12] to minimize the impact of unwanted currents associated with high-resistance state (HRS) of the memory bit-cell (weight =0). If the HRS current ($I_{HRS}$) is not significantly lower compared to the low-resistance state (LRS; weight=1) current ($I_{LRS}$), computational errors can occur. For instance, if $N_H$ cells produce $I_{HRS}$ in an array, the accumulated current $N_H$*$I_{HRS}$ may become larger than $I_{LRS}$ (for large $N_H$ and/or low $I_{LRS}/I_{HRS}$). Thus, this accumulated current is erroneously interpreted as an output ≥ 1 (when ideally, it should be 0). This issue can be countered by using a dummy column with all bit-cells in HRS [11]. The dummy column gets the same inputs as the regular columns. As shown in Fig. 2 (c), the dummy current ($I_{SL,D}$) is subtracted from the SL current ($I_{SL}$) before feeding it to the ADC. This helps in canceling the effect of $N_H$*$I_{HRS}$. If $N_L$ bit-cells in the regular column produce $I_{LRS}$, $I_{SL}$ (neglecting the currents produced by cells with input = 0) is ($N_L I_{LRS}$+$N_H I_{HRS}$), while $I_{SL,D}$ is ($N_L$+$N_H$)$I_{HRS}$. When subtracted, we obtain $N_L$($I_{LRS}$-$I_{HRS}$), which yields $N_L$ as the ADC output. This technique has been shown for spin-based IMC in [11] (since spin-memory have low $I_{LRS}/I_{HRS}$). It is also noteworthy that in the presence of hardware non-idealities such as parasitic resistances, the cancellation of $N_H I_{HRS}$ is not perfect, which we consider in our analysis by including non-idealities with the dummy column.

The crossbar array simulation parameters are listed in Table I. Parasitic resistances viz. driver resistance, wire resistance and via resistance are considered in the array simulations. To perform comprehensive optimization of key device-circuit design knobs considering hardware non-idealities, we use physics-based models to capture the characteristics of each technology, the details of which are provided below.

TABLE I
KEY PARAMETERS FOR CROSSBAR ARRAY

| WL Voltage | 0.7V | BL Voltage | 0.25V |
|---|---|---|---|
| Array Size | 64 × 64 | Driver Resistance | 500Ω |
| Via Resistance | 56Ω[13] | Wire Resistance | 182Ω/µm[13] |
| Bits/Input Signal | 1b | Bits/Device | 1b |

TABLE II
PARAMETERS IN RERAM MODEL

| $I_0$ | $g_0$ | $V_0$ | g |
|---|---|---|---|
| 0.2mA | 0.15nm | 0.35V | 0.34nm-1.09nm |

TABLE III
PARAMETERS IN FEFET MODEL

| FE Thickness [nm] | 7 | 6 | 5 |
|---|---|---|---|
| Saturated Polarization [µC/cm$^2$] | | 30 | |
| Remanent Polarization [µC/cm$^2$] | | 27 | |
| FE Relative Permittivity | 22 | 23.5 | 25 |
| Coercive Electric Field [MV/cm] | 2.4 | 2.525 | 2.65 |

TABLE IV
PARAMETERS IN SOT-MRAM MODEL

| FL Dimension | 30nm × 60nm × 1.8nm |
|---|---|
| Gilbert Damping Constant | 0.007 |
| Spin-Hall Angle | 0.3 |
| Saturation Magnetization | 1257.3 emu/cm$^3$ |
| MgO Thickness | 1.1nm-1.3nm |

### A. Interconnect Modeling

We use 3D COMSOL models for the resistivity and resistance of Cu interconnect with scaled liner/barrier (2nm TaN and 1nm Ta). The model captures surface scattering and grain boundary scattering (details in [13]). *IR* drops on $R_{WIRE}$ are one of the main sources of non-idealities in crossbar arrays [9], especially in deeply scaled technologies. For example, the line metal resistance for M1- M3 considering scaled liner at 7nm technology node is 182Ω/µm [13]. In contrast, at the 45nm technology node, $R_{WIRE}$ is 3.3Ω/µm [14]. More than 50X increase in $R_{WIRE}$ when scaling from 45nm to 7nm node (due to increase in resistivity and reduction in cross-sectional area) is critical, especially for IMC, which is severely affected by *IR* drops.

### B. ReRAM Modeling

We use a compact model of Al-doped $HfO_X$ ReRAM from [15] for our analysis. The ReRAM current is given by

$$I = I_0 * \exp\left(-\frac{g}{g_0}\right) * \sinh\left(\frac{V}{V_0}\right) \tag{1}$$

Here $g$ is the gap length i.e. the distance between the tip of filament and the electrode, $V$ is the applied voltage across the ReRAM cell, $I_0$, $g_0$ and $V_0$ are coefficients. The experimentally calibrated parameters (Table II) are from [15]. To enable selective access (especially programming) in an array, one transistor one resistor (1T1R) bit-cell is used (Fig. 1).

### C. FeFET Modeling

We utilize a compact model of $Hf_{0.5}Zr_{0.5}O_2$-based FeFET in which the ferroelectric layer is modeled with modified Preisach equations [16] and is coupled with a 7nm transistor. The key parameters in the model are shown in Table III which have been obtained by calibrating the model with experiments in [3] and validating using self-consistent phase-field simulation [17] (see [18] for details). The FeFET model can capture the impact of ferroelectric thickness on FeFET-NVM which will be discussed later. FeFET bit-cell is designed with a single FeFET (Fig. 1) by leveraging its self-selecting functionality.

### D. SOT-MRAM Modeling

The dynamics of the magnetic tunnel junction (MTJ) is modeled with non-equilibrium Green's function and Landau-Lifshitz-Gilbert equations [19]. The interaction of the heavy metal with the MTJ is modeled using micromagnetic simulations [4]. Key experimentally calibrated parameters in the SOT-MRAM model are from [20], [21] and listed in Table IV. In our design, tunneling magneto-resistance ratio (TMR) of MTJ ranges from 500% to 600% based on the experiments in [22] and following the work in [4]. Note, for lower TMR values such as 100% -200%, our analysis (and the one in [4]) shows that it is challenging to implement IMC. Two access transistors (for writing and reading/computing each) are used in the SOT-MRAM bit-cell (Fig. 1). The separation of the read and write paths in SOT-MRAMs (unlike spin transfer torque (STT) MRAM) enables independent optimization of the MTJ for IMC, which is important for achieving high IMC robustness (details later).

## IV. EVALUATION METHODOLOGY

To comprehensively evaluate the impact of non-idealities in crossbar arrays on system accuracy, we utilize GENIEX from [2], appropriately modified for our analysis, as discussed subsequently. In the GENIEX framework, extensive HSPICE simulations of synaptic crossbar arrays with various input and weight combinations are performed to obtain the non-ideal current ($I_{non\text{-}ideal}$) and the non-ideality factor (NF) [4] given by

$$NF = \left| \frac{I_{ideal} - I_{non-ideal}}{I_{ideal}} \right| \qquad (2)$$

Here, $I_{ideal}$ is the ideal current and NF quantifies the deviation of $I_{non\text{-}ideal}$ from $I_{ideal}$. NF along with the corresponding input vector and weight matrix are used to train a multi-layer perceptron (MLP). The MLP models a non-ideal crossbar array and predicts NF for any input-weight combination. The MLP is integrated into the Python framework (as in [2]) to evaluate the inference accuracy of ResNet-20 with CIFAR-10 dataset in the presence of device-circuit non-idealities.

While GENIEX in [2] is optimized for D-input configuration, we expand this framework by optimizing the MLPs for G-input. We utilize separate MLPs for each column of a crossbar array (rather than one per array as in [2]), which helps in improving MLP prediction of NF for G-input design. We also incorporate dummy columns in GENIEX. For that, SL current is obtained from HSPICE for the regular ($I_{SL}$) and dummy columns ($I_{SL,D}$). $I_{non\text{-}ideal}$ is calculated as $I_{SL}$-$I_{SL,D}$, which is then used in (2) to obtain NF. Further, we include the effect of process variations (in addition to other hardware non-idealities) in the framework. For that, we consider a Gaussian distribution with mean = 1 and standard deviation=$s$, which is multiplied by the nominal conductance of the bit-cell. Thus, this distribution represents a factor by which the process variations scale the bit-cell conductance. Note that different technologies may exhibit different amounts of variations, which makes the comparison challenging due to the lack of a systematic experimental characterization of variations for emerging technologies. Therefore, we sweep $s$ from 0 to 0.3 in the subsequent sections to illustrate the response of each technology (and its system implications) to process variations.

We also incorporate sense margin (SM) evaluation into our analysis to account for the sensing robustness. This is important because not only does $I_{non\text{-}ideal}$ deviate from $I_{ideal}$, but also the expected MVM output typically needs to be mapped to a range of $I_{non\text{-}ideal}$ (even without the consideration of process variations). This is important while defining the reference levels for the ADCs. To illustrate this point, let us consider the following two scenarios:

- The first scenario is related to the resistance of the bit-cell producing a scalar product of 0. Ideally, the bit-cell resistance should be infinite (zero current); but in reality, this resistance is not only finite, but varies for different input (*In*)-weight (*W*) combinations. Let us define the bit-cell resistances as follows: $R_{OFF}$ for *In*=0, *W* =1, $R_{OFF,H}$ for *In*=0, *W* =0, $R_{HRS}$ for *In*=1, *W*=0, and $R_{ON}$ for *In*=1, *W*=1. $R_{HRS}$, $R_{OFF}$ and $R_{OFF,H}$ correspond to the scalar product of 0 and their values depend on the technology. For instance, let us consider the G-input configuration (Fig. 2(b)). In this case, $R_{OFF,H} >> R_{OFF}$ for 8T-SRAMs (due to stacking effect) and FeFETs (due to higher threshold voltage for weight =0). However, for ReRAMs and SOT-MRAMs, $R_{OFF,H}$ is comparable to $R_{OFF}$ (mainly dictated by the OFF resistance of the access transistor). Furthermore, in FeFETs, ReRAMs and SOT-MRAMs, $R_{HRS} < R_{OFF}$ as the former corresponds to the asserted state of the bit-cell, while the latter is for the de-asserted state. On the other hand, for 8T-SRAMs, $R_{HRS}$ and $R_{OFF}$ are comparable. These comparisons are quantified later; here, we would like to emphasize the point that different *In*-*W* combinations for the same scalar product yield different bit-cell resistances and currents. As a result, the IMC output can also be different. More importantly, the difference between these currents is dependent on the technology. It is worth mentioning that including a dummy column [11] in each crossbar array can partially compensate the impact of finite $R_{HRS}$, as we discussed earlier.
- The second scenario is associated with the location of the bit-cell in an array producing a non-zero scalar product. This is important due to the $R_{WIRE}$ and is especially critical for scaled technologies in which $R_{WIRE}$ is large. To illustrate this, let us consider two conditions corresponding to the dot product = 1: (i) input = [1,0…0], weight = [1,0…0] and (ii) input = [0,0…1], weight = [0,0…1]. In case (i), the first bit-cell (farthest from the sensing circuit) produces the scalar product of 1, while in case (ii), the same is true for the last bit-cell (closest to the sensing circuit). Thus, the effective $R_{WIRE}$ seen by the source of the ON bit-cell is higher for case (i), while that seen by the drain is higher for case (ii). Depending on the relative sensitivity of $R_{ON}$ to the voltage on the D-terminal ($V_D$) and S-terminal ($V_S$) of the bit-cell, different SL currents are obtained for case (i) and (ii). Similarly, for other input-weight combinations, the currents could be different, although they represent the same dot product output.

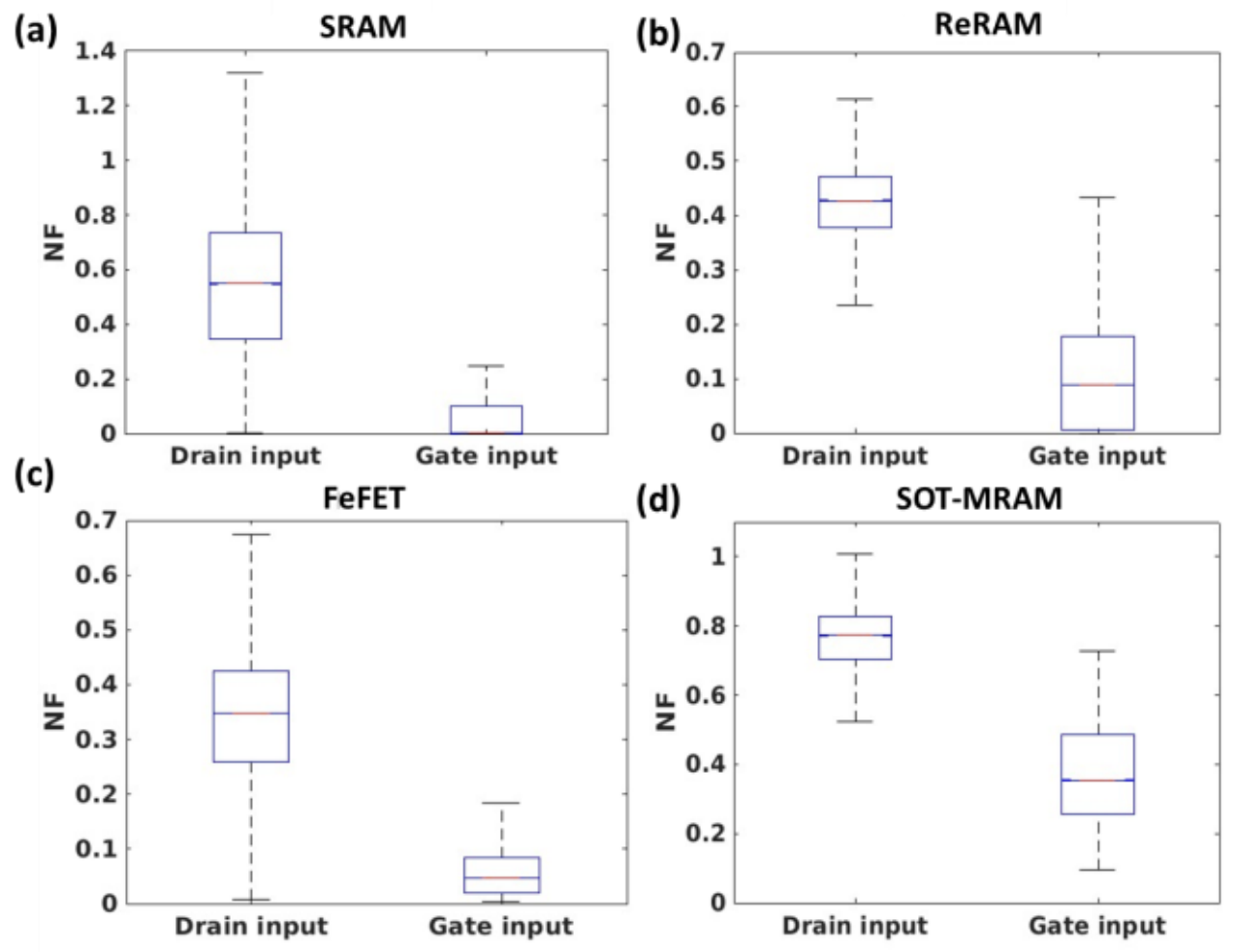

Fig. 3. Box plot of NF showing that NF of D-input configuration is more than NF of G-input configuration for (a) SRAMs (b) ReRAMs (c) FeFETs (d) SOT-MRAMs.

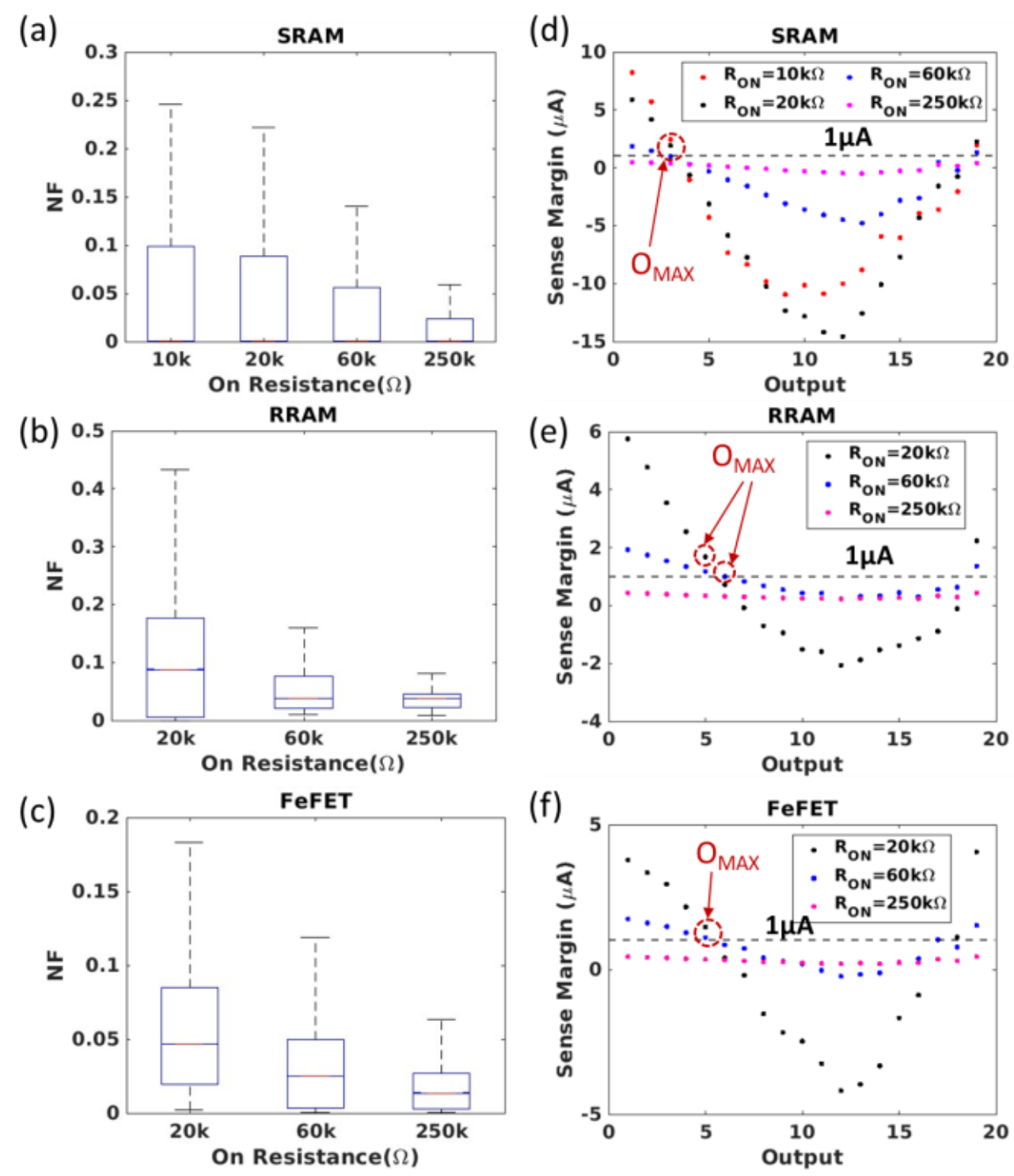

Fig. 4. Box plot of NF with various $R_{ON}$ for (a) SRAMs (b) ReRAMs (c) FeFEsT. SM versus output with various $R_{ON}$ for (d) SRAMs (e) ReRAMs (f) FeFETs.

Through extensive HSPICE circuit simulations (for various near-extreme input-weight combinations), we obtain the minimum and maximum currents representing each output. SM between output x and output x-1 is defined as:

$$SM_X = \frac{I_{X,MIN} - I_{X-1,MAX}}{2} \tag{3}$$

$I_{X,MIN}$ is the minimum current when output is $x$, and $I_{X-1,MAX}$ is the maximum current when output is $x$-1. We obtain SM for different outputs to evaluate the sensing robustness. As a part of the subsequent design optimizations, SM analysis is utilized in conjunction with the NF analysis to establish the optimal design choice for minimizing the impact of non-idealities on IMC robustness. Note that both NF and SM-based analyses have their own merits. NF represents a statistical analysis based on the workload-specific input and weight distributions but does not consider the ADC sensing limitations. On the other hand, SM-based analysis is a more comprehensive array evaluation method accounting for array non-idealities and practical ADC capabilities, but is based on the (near) worst-case scenario and does not capture the workload characteristics. Hence, we utilize both to study the optimization strategies.

## V. Device-Circuit Co-Optimizations

In this section, we perform technology-agnostic and technology-specific optimizations considering the cross-layer device-circuit interactions to establish the optimal design metrics for each technology in the presence of non-idealities.

### A. D-Input versus G-Input

Fig. 3 compares the NF of the D-input and G-input configurations for different technologies at the 7nm node. NF for D-input is much larger than G-input, indicating the higher effect of non-idealities in the former. This is because the current-carrying terminals in D-input (BL and SL) run orthogonally, and thus, the current through any bit-cell is directly dependent on the conductance of bit-cells in its row *and* column, and, indirectly, on all the bit-cells in the array (through sneak current paths). On the other hand, for G-input, the current-carrying terminals are routed vertically along the column, and hence, the current in a bit-cell is dependent on the bit-cells *only* in that column. This reduction of data dependency for G-input lowers the impact of $R_{WIRE}$ as well as the range of currents for each output, yielding lower NF. At the 7nm node, the impact of $R_{WIRE}$ is large (as noted before). Hence the non-idealities for D-input increase immensely. To manage the non-idealities at the 7nm node and achieve acceptable level of accuracies, we use G-input.

### B. ON Resistance Optimization

$R_{ON}$ optimization is critical for IMC robustness because of its direct impact on the output current. As $R_{ON}$ reduces, the output current increases, which has two opposing effects. On one hand, it leads to an increase in non-idealities associated with *IR* drop on parasitic resistances, which increases NF [4]. On the other hand, the SM improves due to higher current, which reduces the sensing errors. Thus, it is important to design $R_{ON}$ considering the trade-off between NF and SM.

In 8T-SRAMs, $R_{ON}$ is optimized by applying different bias voltage ($V_{BIAS}$) to gates of *M1* and *M2* (Fig. 1 (a)). In our study, we sweep $V_{BIAS}$ from 0.7V (=$V_{DD}$) to 0.45V tuning $R_{ON}$ from 10kΩ to 250kΩ. As $R_{ON}$ increases, NF decreases (Fig. 5(a)), as expected. To understand the effect of $R_{ON}$ on SM, let us look at Fig. 4(d). We define $O_{MAX}$ as the maximum output that can be sensed in a robust fashion by ADC (considering SM > 1μA). SM for $R_{ON}$ = 250kΩ is always below 1μA which indicates poor sensing robustness while $O_{MAX}$ for $R_{ON}$ = 10 kΩ, 20kΩ, and 60kΩ are similar. Although SM for some high outputs can be negative, these represent the near-worst-case inputs/weight combinations, which are unlikely to occur for a DNN workload. Moreover, the occurrence of high output values is rare due to sparsity [23]. Therefore, some sensing errors may be acceptable. For SRAMs, we design $R_{ON}$ to be around 60kΩ ($V_{BIAS} \approx 0.52$V) based on NF and SM.

$R_{ON}$ of ReRAMs can be tuned by changing the gap length ($g$) which ranges from 0.34nm to 1.09nm, as per [15]. The maximum $g$ (=1.09nm) corresponds to the HRS of ReRAMs. The minimum resistance of ReRAMs is 20 kΩ when $g$ is 0.34nm. We sweep $R_{ON}$ of ReRAMs from 20 kΩ to 250 kΩ by changing $g$ from 0.34nm to 0.75nm. Based on Fig. 4, we optimize $R_{ON}$ to be around 60kΩ ($g$ = 0.53nm) considering NF and SM.

For $R_{ON}$ optimization in FeFETs, we tune the set voltages to program an optimal threshold voltage of the FeFETs during the set state. For that, reset voltage ($V_{RESET}$ = -5$V$) is first applied to the gate of FeFETs to reset the whole array (global reset [24]). Then set voltage ($V_{SET}$) is applied to the gate of FeFETs to program weight = 1 selectively. We employ 'V/2' biasing scheme for the unselected/half-selected cells as in [25]. In our analysis, we sweep $R_{ON}$ of FeFETs from 20 kΩ to 250 kΩ by tuning $V_{SET}$ (Fig. 4). We find that the $R_{ON}$ around 60kΩ ($V_{SET}$ = 2.2V) is near-optimal for FeFETs considering the NF and SM analysis.

For SOT-MRAMs, $R_{ON}$ can be optimized by tuning the tunneling oxide (MgO) thickness or gate voltage of the read access transistor. However, that also changes $R_{HRS}$, which is quite low in the MTJs due to their low TMR. Therefore, unlike the other three technologies, SOT-MRAMs need to consider both $R_{ON}$ and $R_{HRS}$ at the same time, which we will discuss in sub-section D.

Further, besides $R_{ON}$, the optimization of $R_{OFF}$ and $R_{HRS}$ can also be important. For SRAMs, $R_{HRS}$ and $R_{OFF}$ are inherently very high in SRAMs, as in both cases, one of the transistors in the read port is OFF. For ReRAMs, we maximize $R_{HRS}$ by using maximum gap length (as noted earlier). $R_{OFF}$ for ReRAMs is high as the access transistor is OFF. For FeFETs, due to the absence of an access transistor, we need to consider other design knobs to optimize $R_{HRS}$ and $R_{OFF}$ (Fig. 1). In this work, we focus on FE thickness ($T_{FE}$) optimization for this purpose, as discussed next.

### C. Ferroelectric Thickness ($T_{FE}$) Optimization in FeFETs

$T_{FE}$ is a key design parameter in FeFETs and plays a key role in IMC robustness [18]. In our analysis, we fix the optimized $R_{ON}$ value (discussed before) for different $T_{FE}$ by tuning $V_{SET}$. It may be noted that $T_{FE}$ = 10nm used in several experimental works [3], [25] led to unacceptably high short channel effects for the 7nm technology node and therefore, is not included in our analysis. We observe $T_{FE}$ = 7nm to be the maximum value that provided reasonably strong gate control with OFF current ~ 60nA. We observe that NF increases with $T_{FE}$ scaling from 7nm to 5nm (Fig. 5 (a)). We can understand this trend from $I_{DS}$-$V_{GS}$ characteristics of FeFETs for different $T_{FE}$ [18]. As shown in Fig. 5 (b), $I_{HRS}$ (corresponding to $In$=1, $W$=0) increases as $T_{FE}$ scales due to the reduction in the memory window. This leads to the increase in NF as $T_{FE}$ scales from 7nm to 5nm. Thus, $T_{FE}$ around 7nm is the optimal design point and used for subsequent analysis. It may be noted that while the dummy column reduces the impact of $I_{HRS}$ (as discussed before), its alleviation is only partial due to parasitic resistances. Therefore, low $I_{HRS}$ yields low NF, even when the dummy column is considered (as in our analysis).

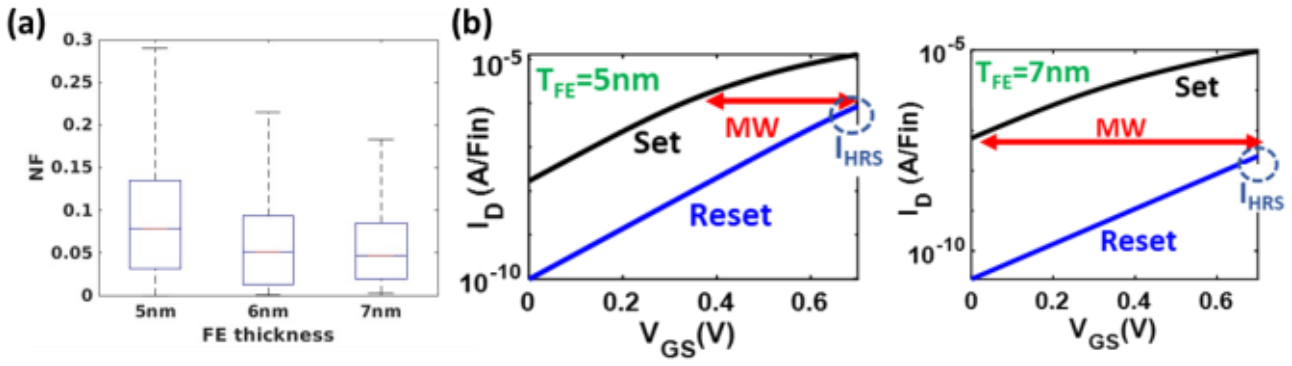

Fig. 5. (a) Box plot of NF for FeFETs with various $T_{FE}$. FeFETs $I_{DS}$-$V_{GS}$ characteristics at $V_{DS}$ = 0.25V for $T_{FE}$ = 5nm/7nm.

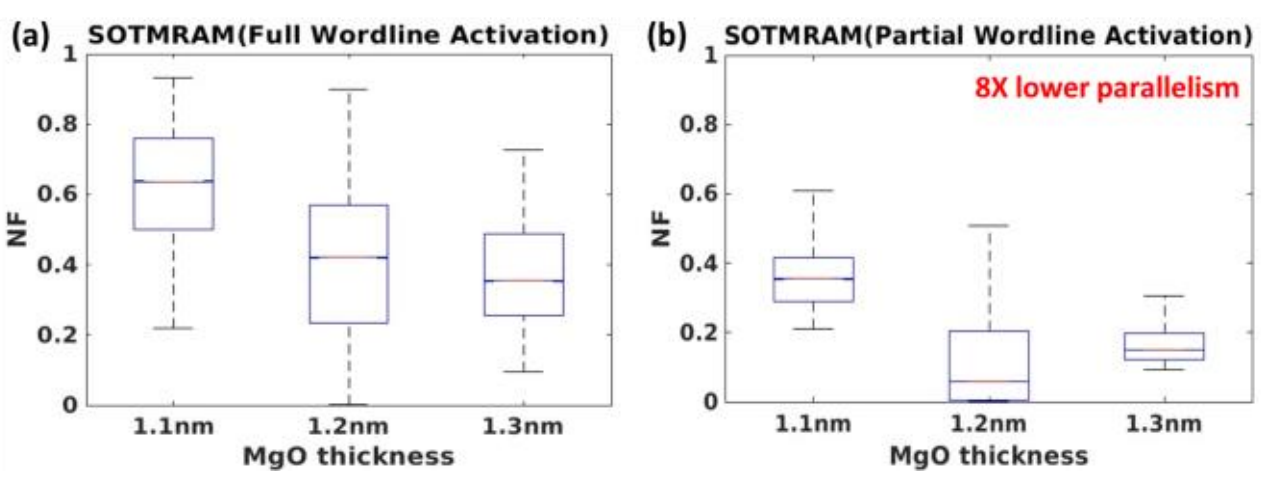

Fig. 6. (a) Box plot of NF for SOT-MRAMs with various $T_{MgO}$ with full wordline activation. (b) Box plot of NF for SOT-MRAMs with various $T_{MgO}$ with partial wordline activation.

TABLE V
RESISTANCE OF SOT-MRAM FOR DIFFERENT MGO THICKNESS

| MgO Thickness (nm) | 1.1 | 1.2 | 1.3 |
|---|---|---|---|
| $R_{ON}$ (kΩ) | 8 | 12 | 20 |
| $R_{HRS}$(kΩ) | 28 | 52 | 100 |
| $R_{HRS}/R_{ON}$ | 3.5 | 4.3 | 5 |

### D. MgO Thickness ($T_{MgO}$) Optimization in SOT-MRAMs

We sweep $T_{MgO}$ from 1.1nm to 1.3nm and tune $R_{ON}$ and $R_{HRS}$ of SOT-MRAMs as shown in Table V. As $T_{MgO}$ increases, NF decreases (Fig. 6(a)) due to lower bit-cell currents leading to less $IR$ drop on the parasitic resistances. The $R_{HRS}/R_{ON}$ ratio also increases as $T_{MgO}$ increases which further reduces the non-idealities. Increasing $T_{MgO}$ beyond 1.3 nm leads to low read currents and SM, even for conventional read (i.e. single row assertion). Thus, $T_{MgO}$ = 1.3nm is the optimal choice.

However, even for SOT-MRAMs with $T_{MgO}$ =1.3nm, NF is still high due to the low $R_{HRS}/R_{ON}$ ratio (or low TMR) of the MTJ. To reduce NF, we use partial WL activations (PWA) in which 8 rows in a 64 × 64 crossbar array are activated simultaneously in one cycle. Hence, 8 cycles are needed for the entire matrix-vector multiplication of the array. Compared with full WL activation (FWA) in which all the 64 rows in a 64 × 64 crossbar array are activated simultaneously, the NF of PWA decreases (Fig. 6 (b)) due to lower currents in the column.

Therefore, we use $T_{MgO}$ =1.3nm with PWA (8 rows activated in a single cycle) as the design choice for SOT-MRAM for subsequent analysis. It may be noted that for the other three technologies, FWA is used to maximize parallelism, while still achieving low NF and high SM.

## VI. INFERENCE ACCURACY AND TECHNOLOGY COMPARISON

Let us now compare the inference accuracy of different technologies for ResNet-20 on CIFAR-10 dataset. Note that the key design parameters of each technology have been optimized according to the analysis in the last section for fair comparison. Fig. 7 shows the accuracy of each technology considering variations with varying standard deviations: $s$=0, 0.1, 0.2 and 0.3. Note that $s$=0 corresponds to the nominal case (i.e. without variations). In the nominal case, the inference accuracies for FeFETs and ReRAMs are comparable (with FeFETs showing mildly larger accuracy). FeFETs and ReRAMs exhibit the highest accuracy followed by SRAMs while that of SOT-MRAMs is the lowest. When device variations are considered (s=0.1,0.2 and 0.3), the degradation of system accuracy is notable for SOT-MRAMs. FeFETs-based DNNs show the least impact of variations on system accuracy followed by ReRAMs and SRAMs. The trend of accuracy for different technologies can be explained as follows.

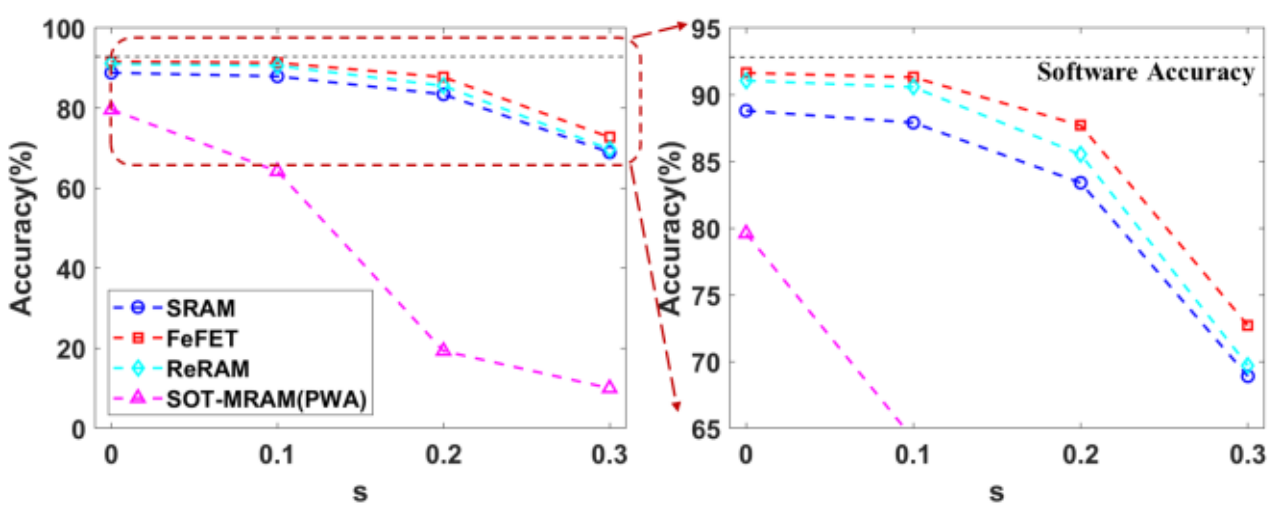

Fig. 7. Accuracy comparison of various technologies considering variations with varying standard deviations (*s*=0/0.1/0.2/0.3) for CIFAR-10 dataset on ResNet-20. The accuracy of the software baseline is 92.8%.

TABLE VI
RESISTANCE COMPARISON FOR DIFFERENT TECHNOLOGIES

| | $R_{ON}$(kΩ) | $R_{HRS}$(kΩ) | $R_{OFF}$(kΩ) | $R_{OFF,H}$(kΩ) |
|---|---|---|---|---|
| SRAM | 60 | $2.1*10^8$ | $1.5*10^8$ | $4.5*10^8$ |
| FeFET | 60 | $4*10^3$ | $2.3*10^4$ | $4.6*10^6$ |
| ReRAM | 60 | $2.3*10^3$ | $2.1*10^5$ | $2.1*10^5$ |
| SOT-MRAM | 20 | 100 | $2.1*10^5$ | $2.1*10^5$ |

In the G-input configuration, the current-carrying wires are routed along the column and hence, $R_{WIRE}$ is dependent on the vertical height of the bit-cell. As illustrated in Fig. 1 (b), the layouts of bit-cells of different technologies have been optimized to reduce the vertical height as much as possible so as to align with the needs of G-input. Now, the self-selecting functionality of FeFETs leads to the sharing of both source and drain contacts of FeFETs along the column. Hence, the vertical height of the FeFET bit-cell is 1GP (where GP is the gate pitch). In the ReRAM bit-cell, only one contact of the access transistor can be shared along the column, leading to layout height = 1.5GP. The vertical height is 2GP for the 8T-SRAM bit-cell and SOT-MRAM bit-cell. Thus, $R_{WIRE}$ is the least for FeFETs due to their smallest layout height.

In addition, $R_{ON}$, $R_{HRS}$ $R_{OFF}$ and $R_{OFF,H}$ of different technologies are crucial parameters, which are compared in Table VI. Since SOT-MRAMs have the lowest $R_{ON}$ and $R_{HRS}$, *IR* drop for SOT-MRAMs is the most detrimental to accuracy. The low $R_{HRS}/R_{ON}$ ratio in SOT-MRAMs also aggravates the non-idealities. As a result, SOT-MRAMs exhibit the lowest accuracy (despite the PWA scheme). $R_{ON}$ of SRAMs, FeFETs and ReRAMs are similar (due to the aforementioned optimizations); however, their $R_{HRS}$ and $R_{OFF}$ are quite different. SRAMs exhibit the highest $R_{HRS}$ (as the read port is OFF), followed by ReRAMs and FeFETs. $R_{OFF}$ of SRAMs is also the highest (due to low standby power transistors used (see Section III). FeFETs and ReRAMs exhibit reasonably high $R_{HRS}$ and $R_{OFF}$ (although not as high as SRAMs).

Another critical factor affecting the accuracy is the sensitivity of bit-cell current for each technology to the *IR* drops. As noted earlier, *IR* drop on the $R_{WIRE}$ leads to data-dependent variations in $V_D$ and $V_S$. Specifically, $V_D$ reduces (from its ideal value) and $V_S$ increases due to the *IR* drop, and the extent of this reduction/increase depends on the input-weight combinations. This results in variations in the bit-cell currents, which increases the range of NF and reduces IMC robustness. If we examine the bit-cells in Fig. 1 (a), it can be observed that increasing $V_S$ reduces (i) $V_{GS}$ of M2 in SRAMs, $V_{GS}$ of access transistors in ReRAMs and SOT-MRAMs, and $V_{GS}$ of FeFETs, and (ii) voltage across the bit-cell ($V_{DS}$). The first effect leads to source-degeneration of the transistors. On the other hand, reducing $V_D$ just leads to lower $V_{DS}$. Our analysis shows that the sensitivity of bit-cell current to $V_{DS}$ is only mildly different in the four technologies (for the design points used in this work). In contrast, the sensitivity of bit-cell to $V_{GS}$ is quite different. In ReRAMs and SOT-MRAMs (and for that matter, any 1T1R bit-cell), the design optimizations typically ensure that the access transistor resistance is much smaller than the memory device resistance to achieve sufficient distinguishability. Hence, the effect of source degeneration is minimal, leading to the small sensitivity of bit-cell current to $V_S$ in ReRAMs and SOT-MRAMs. In this respect, the former is better than the latter due to a higher resistance of ReRAMs device compared to MTJs (as can be inferred from the $R_{ON}$ comparison in Table IV), which reduces the relative resistance of the access transistor. However, in SRAMs and FeFETs, the source degeneration strongly reduces the bit-cell current, leading to a much larger sensitivity of bit-cell current to $V_S$. Thus, ReRAMs exhibit the lowest sensitivity of bit-cell current to *IR* drops.

Now, let us put together these three factors viz. layout height, HRS/OFF resistances and sensitivity of bit-cell currents to the *IR* drops. FeFETs combine the benefits of compactness (low *IR* drop), reasonably high $R_{OFF}$ and $R_{HRS}$ (achieved via $T_{FE}$ optimization) and optimized $R_{ON}$ to achieve the highest accuracy amongst the four technologies. In the nominal case, the performance of ReRAMs is comparable to FeFETs. On the one hand, ReRAMs are less compact than FeFETs leading to higher $R_{WIRE}$. On the other hand, since bit-cell current in ReRAMs is less sensitive to *IR* drops, the effect of high $R_{WIRE}$ is offset. Moreover, $R_{ON}$ optimization and high $R_{OFF}$ and $R_{HRS}$ help in improving the accuracy. Despite offering the highest $R_{HRS}$, SRAMs suffer from large memory footprints, leading to high $R_{WIRE}$. Moreover, the sensitivity of their bit-cell current to *IR* drops is also high, leading to some degradation in accuracy compared to FeFETs and ReRAMs. SOT-MRAMs suffer from low $R_{HRS}/R_{ON}$ ratio, low $R_{ON}$ and large bit-cell area, leading to the least accuracy amongst the four technologies. The appealing attributes of FeFETs lead to the highest accuracy especially when variations are considered. The effect of device variations on DNN accuracy is aggravated due to device-circuit non-idealities. As FeFET-based crossbar arrays exhibit the lowest non-idealities, their sensitivity to variations is also lower compared to the other technologies.

## VII. CONCLUSION

This work performs a comparative technology evaluation and provides insights into the design of synaptic crossbar arrays in the presence of non-idealities at the scaled (7nm) technology node. Based on a cross-layer design flow, we evaluate device-circuit non-idealities and system accuracy implications of synaptic crossbar arrays based on 8T-SRAMs, ReRAMs, FeFETs and SOT-MRAMs. After an extensive co-optimization of design parameters considering non-idealities, we compare different technologies in terms of the inference accuracy for ResNet-20 on CIFAR-10 dataset. Our results show that under nominal conditions, FeFETs-based DNNs achieve the highest accuracy followed closely by ReRAMs. Further, FeFETs lead to the largest resilience to process variations. These system-level benefits are achieved by virtue of high $R_{OFF}/R_{HRS}$ and optimized $R_{ON}$ of FeFET bit-cells as well as their low area which reduces the impact of $R_{WIRE}$.

## Acknowledgment

We would like to thank X. Chen, A. Malhotra, K. Cho, T, Sharma and K. Roy at Purdue University for various useful discussions.

## References

[1] A. Jaiswal, I. Chakraborty, A. Agrawal, and K. Roy, "8T SRAM cell as a multibit dot-product engine for beyond von Neumann computing," *IEEE Transactions on Very Large Scale Integration (VLSI) Systems*, vol. 27, no. 11, pp. 2556–2567, Nov. 2019, doi: 10.1109/TVLSI.2019.2929245.

[2] I. Chakraborty, M. Fayez Ali, D. Eun Kim, A. Ankit, and K. Roy, "GENIEx: A generalized approach to emulating non-ideality in memristive xbars using neural networks," presented at the Proceedings - Design Automation Conference, 2020. doi: 10.1109/DAC18072.2020.9218688.

[3] K. Ni, B. Grisafe, W. Chakraborty, A. K. Saha, S. Dutta, M. Jerry, J. A. Smith, S. Gupta, and S. Datta, "In-memory computing primitive for sensor data fusion in 28 nm HKMG FeFET technology," presented at the Technical Digest - International Electron Devices Meeting, IEDM, 2019. doi: 10.1109/IEDM.2018.8614527.

[4] T. Sharma, C. Wang, A. Agrawal, and K. Roy, "Enabling robust SOT-MTJ crossbars for machine learning using sparsity-aware device-circuit co-design," presented at the Proceedings of the International Symposium on Low Power Electronics and Design, 2021. doi: 10.1109/ISLPED52811.2021.9502492.

[5] K. He, I. Chakraborty, C. Wang, and K. Roy, "Design space and memory technology co-exploration for in-memory computing based machine learning accelerators," presented at the IEEE/ACM International Conference on Computer-Aided Design, Digest of Technical Papers, ICCAD, 2022. doi: 10.1145/3508352.3549453.

[6] P. Y. Chen, X. Peng, and S. Yu, "NeuroSim: A circuit-level macro model for benchmarking neuro-inspired architectures in online learning," *IEEE Transactions on Computer-Aided Design of Integrated Circuits and Systems*, vol. 37, no. 12, 2018, doi: 10.1109/TCAD.2018.2789723.

[7] S. Jain and A. Raghunathan, "CxDNN: Hardware-software compensation methods for deep neural networks on resistive crossbar systems," *ACM Transactions on Embedded Computing Systems*, vol. 18, no. 6, 2019, doi: 10.1145/3362035.

[8] S. K. Roy, A. Patil, and N. R. Shanbhag, "Fundamental limits on the computational accuracy of resistive crossbar-based in-memory architectures," presented at the Proceedings - IEEE International Symposium on Circuits and Systems, 2022. doi: 10.1109/ISCAS48785.2022.9937336.

[9] I. Chakraborty, M. Ali, A. Ankit, S. Jain, S. Roy, S. Sridharan, A. Agrawal, A. Raghunathan, and K. Roy, "Resistive crossbars as approximate hardware building blocks for machine learning: Opportunities and challenges," presented at the Proceedings of the IEEE, 2020. doi: 10.1109/JPROC.2020.3003007.

[10] L. Clark, V. Vashishtha, L. Shifren, A. Gujja, S. Sinha, B. Cline, C. Ramamurthy, and G. Yeric, "ASAP7: A 7-nm finFET predictive process design kit," *Microelectronics Journal*, vol. 53, 2016, doi: 10.1016/j.mejo.2016.04.006.

[11] K. Cho, X. Fong, and S. K. Gupta, "Exchange-coupling-enabled electrical-isolation of compute and programming paths in valley-spin Hall effect based spintronic device for neuromorphic applications," presented at the Device Research Conference - Conference Digest, DRC, 2021. doi: 10.1109/DRC52342.2021.9467139.

[12] A. Shafiee, A. Nag, N. Muralimanohar, R. Balasubramonian, J. P. Strachan, M. Hu, R. S. Williams, and V. Srikumar, "ISAAC: A convolutional neural network accelerator with in-situ analog arithmetic in crossbars," presented at the Proceedings - 2016 43rd International Symposium on Computer Architecture, ISCA 2016, 2016. doi: 10.1109/ISCA.2016.12.

[13] X. Chen, C. L. Lo, M. C. Johnson, Z. Chen, and S. K. Gupta, "Modeling and circuit analysis of interconnects with TaS2 barrier/liner," presented at the Device Research Conference - Conference Digest, DRC, 2021. doi: 10.1109/DRC52342.2021.9467160.

[14] P. Moon, V. Chikarmane, K. Fischer, R. Grover, T. A. Ibrahim, and K. J. Lee, "Process and electrical results for the on-die interconnect stack for Intel's 45 nm process generation," *Intel Technol. J.*, vol. 12, pp. 87–92, Jan. 2008.

[15] Z. Jiang, Y. Wu, S. Yu, L. Yang, K. Song, Z. Karim, and H.-S. P. Wong, "A Compact model for metal-oxide resistive random access memory with experiment verification," *IEEE Transactions on Electron Devices*, vol. 63, no. 5, pp. 1884–1892, May 2016, doi: 10.1109/TED.2016.2545412.

[16] A. K. Saha and S. K. Gupta, "Modeling and comparative analysis of hysteretic ferroelectric and anti-ferroelectric FETs," presented at the Device Research Conference - Conference Digest, DRC, 2018. doi: 10.1109/DRC.2018.8442136.

[17] A. K. Saha, M. Si, K. Ni, S. Datta, P. D. Ye, and S. K. Gupta, "Ferroelectric thickness dependent domain interactions in FEFETs for memory and logic: A phase-field model based analysis," presented at the Technical Digest - International Electron Devices Meeting, IEDM, 2020. doi: 10.1109/IEDM13553.2020.9372099.

[18] C. Wang, J. Victor, A. K. Saha, X. Chen, M. Si, T. Sharma, K. Roy, P. D. Ye, and S. K. Gupta, "FeFET-based synaptic cross-bar arrays for deep neural networks: Impact of ferroelectric thickness on device-circuit non-idealities and system accuracy," presented at the 2023 Device Research Conference (DRC), IEEE, Jun. 2023, pp. 1–2. doi: 10.1109/DRC58590.2023.10187042.

[19] X. Fong, S. K. Gupta, N. N. Mojumder, S. H. Choday, C. Augustine, and K. Roy, "KNACK: A hybrid spin-charge mixed-mode simulator for evaluating different genres of spin-transfer torque MRAM bit-cells," presented at the International Conference on Simulation of Semiconductor Processes and Devices, SISPAD, 2011. doi: 10.1109/SISPAD.2011.6035047.

[20] S. Ikeda, K. Miura, H. Yamamoto, K. Mizunuma, H. D. Gan, M. Endo, S. Kanai, J. Hayakawa, F. Matsukura and H. Ohno, "A perpendicular-anisotropy CoFeB-MgO magnetic tunnel junction," *Nature Materials*, vol. 9, no. 9, 2010, doi: 10.1038/nmat2804.

[21] C. F. Pai, L. Liu, Y. Li, H. W. Tseng, D. C. Ralph, and R. A. Buhrman, "Spin transfer torque devices utilizing the giant spin Hall effect of tungsten," *Applied Physics Letters*, vol. 101, no. 12, 2012, doi: 10.1063/1.4753947.

[22] S. Ikeda,1, J. Hayakawa, Y. Ashizawa, Y. M. Lee, K. Miura, H. Hasegawa, M. Tsunoda, F. Matsukura, and H. Ohno1, "Tunnel magnetoresistance of 604% at 300K by suppression of Ta diffusion in CoFeB/MgO/CoFeB pseudo-spin-valves annealed at high temperature," *Applied Physics Letters*, vol. 93, no. 8, p. 082508, Aug. 2008, doi: 10.1063/1.2976435.

[23] S. K. Thirumala, S. Jain, S. K. Gupta, and A. Raghunathan, "Ternary compute-enabled memory using ferroelectric transistors for accelerating deep neural networks," presented at the Proceedings of the 2020 Design, Automation and Test in Europe Conference and Exhibition, DATE 2020, 2020. doi: 10.23919/DATE48585.2020.9116495.

[24] M. Trentzsch, S. Flachowsky, R. Richter, J. Paul, B. Reimer, D. Utess, S. Jansen, H. Mulaosmanovic, S. Müller, S. Slesazeck, J. Ocker, M. Noack, J. Müller, P. Polakowski, J. Schreiter, S. Beyer, T. Mikolajick, and B. Rice, "A 28nm HKMG super low power embedded NVM technology based on ferroelectric FETs," presented at the Technical Digest - International Electron Devices Meeting, IEDM, 2017. doi: 10.1109/IEDM.2016.7838397.

[25] K. Ni, X. Li, J. A. Smith, M. Jerry, and S. Datta, "Write disturb in ferroelectric FETs and its implication for 1T-FeFET and memory arrays," *IEEE Electron Device Letters*, vol. 39, no. 11, 2018, doi: 10.1109/LED.2018.2872347.